\documentclass[
 reprint,twocolumn,
 amsmath,amssymb,
 aps,
]{revtex4-1}

\usepackage{epsfig,bm,epsf,graphics}
\usepackage{graphicx,subfigure}
\usepackage{dcolumn}
\usepackage{bm}
\usepackage{xcolor}
\DeclareMathOperator{\para}{\!\tiny{\mbox{$/ \!\! /$}}}

\begin{document}
\preprint{APS/123-QED}

\title{Vortex Structure in Magnetic Nanodots: Dipolar Interaction,
Mobile Spin Model, Phase Transition and Melting}

\author{Aur\'elien Bailly-Reyre$^{1}$
and H. T. Diep\footnote{corresponding author}$^2$}
\address{%
$^1$ \quad LPNHE - Laboratoire de Physique Nucl\'eaire et de Hautes \'Energies,
Sorbonne Universit\'e, Universit\'e de Paris, CNRS/IN2P3, 4 Place
Jussieu, 75005 Paris, France ; abaillyr@lpnhe.in2p3.fr\\
$^{2}$ \quad Laboratoire de Physique Th\'eorique et Mod\'elisation,
CY Cergy Paris Universit\'e\footnote{Formerly, University of Cergy-Pontoise}, CNRS,
UMR 8089, 2 Avenue Adolphe Chauvin, 95302 Cergy-Pontoise, France; diep@cyu.fr, corresponding author}
\date{\today}

\begin{abstract}
We study in this article properties of a nanodot embedded in a support by Monte Carlo simulation.  The nanodot is a piece of simple cubic lattice where each site is occupied by a mobile Heisenberg spin which can move from one lattice site to another under the effect of the temperature and its interaction with neighbors.  We take into account a short-range exchange interaction between spins and a long-range dipolar interaction.  We show that the ground-state configuration is a vortex around the dot central axis: the spins on the dot boundary lie  in the $xy$ plane but go out of plane with a net perpendicular magnetization at the dot center. Possible applications are discussed. Finite-temperature properties are studied. We show the characteristics of the surface melting and determine the energy, the diffusion coefficient and the layer magnetizations as functions of temperature.
\vspace{0.5cm}
\begin{description}
\item[PACS numbers: 75.10.-b ; 75.10.Hk ; 64.60.Cn ]
\end{description}
\end{abstract}

\pacs{Valid PACS appear here}
\maketitle


\section{Introduction}

In finite systems such as magnetic thin films and nanodots, spin configurations are often non uniform near the surface.  The ground-state (GS) structure results from the competition between interactions in the system. The combination of the frustration \cite{DiepFSS} resulting from competing interactions and the boundary effects in finite systems gives rise to unexpected phenomena \cite{Diep2013}.  Among the competing forces, let us focus on the dipolar interaction which favors an in-plane non uniform spin configuration in flat and small samples, and the exchange interaction which tends to align spins in parallel configuration.  In addition, one can have the presence of a perpendicular anisotropy is known to arise with a large magnitude in ultrathin films \cite{zangwill,bland-heinrich}.  Note that, in thin films with Heisenberg and Potts models, the competition between the dipolar interaction, the exchange interaction and  the perpendicular anisotropy causes
a spin re-orientation transition at a finite temperature \cite{Santa2000,Hoang2013}. There has been a great number of other works treating the dipolar interaction in the the presence of a perpendicular anisotropy in 2D monolayers and thin films. All of them found various ground states (GS) such as in-plane, out-of-plane and non-uniform strip-domain configurations. Let us mention a few of them: in Ref. \onlinecite{Yafet} an analytical calculation has been performed at zero temperature ($T$) to find GS by varying the uniaxial surface anisotropy in a monolayer and in thin films (i. e. infinite lateral dimension). In Refs. \onlinecite{Moschel,MacIsaac,Vedmedenko}, Monte Carlo (MC) simulations have been carried out for monolayers and thin films at finite $T$ where re-orientation phase transitions have been found. Except in Ref. \onlinecite{Vedmedenko} where all transitions are of second order, the two other works found first- and second-order transitions depending of the ratio of perpendicular anisotropy to dipolar strength.  In Ref. \onlinecite{Maziewski}, the authors used micromagnetic simulations to calculate the $T=0$ configurations for wires and disks. They did not consider finite-$T$ behaviors. In Ref. \cite{Xiao}, a spin-reorientation has been experimentally observed in Nd$_2$Fe$_{14}$B by using Lorentz transmission electron microscopy at variable temperatures and magnetic fields. It was shown that skyrmions are created around the spin-reorientation  temperature.    The absence of works dealing with ultrafine dots at finite $T$ using a mobile spin model has motivated the present work.

In this paper, we focus on the case of a small magnetic nanodot with mobile Heisenberg spins.   Various GS configurations have been observed in such nanosystems \cite{RochaJAP107}, depending on the size of the sample, the ratio between the exchange and dipolar interactions, and the type of the lattice. Systems in which the core vortex structure occurs hold much promise from the commercial point of view; the occurrence of this structure has already been demonstrated experimentally \cite{Shinjo289,Raabe88,Wachowiak298} by different imaging techniques. A major advantage of core vortex structures is the central region (core) of nonzero perpendicular magnetization, the polarization of which is stable at room temperature as shown by Shinjo \emph{et al.} \cite{Shinjo289}. Interestingly, core magnetization reversal \cite{Kikuchi90, Xiao102} can be realized in two ways, by applying a strong magnetic field \emph{perpendicular} to the surface of the sample, or a short pulse of magnetic field \emph{parallel} to it. This property of magnetic nanodots opens the door to their application in magnetoresistive random access memory (MRAM).

The current development of a technology that allows to obtain samples with a very small dimension \cite{Broeder60,Kurt108,Hodumi90} has inspired us to investigate, with the use of MC simulations \cite{Landau09,Brooks11}, the behavior of the core vortex structure, so interesting from the point of view of applications, under the impact of the temperature $T$ using a mobile spin model.  The mobility of the spins with increasing $T$ gives the opportunity to investigate the melting of these nanodots.

The GS structure found for a nanodot can be considered as a single skyrmion \cite{Skyrme}. There are several mechanisms and interactions leading to the appearance of skyrmions in various kinds of matter. The most popular one is certainly the Dzyaloshinskii-Moriya (DM) interaction which was initially proposed to explain the weak ferromagnetism observed in antiferromagnetic Mn compounds.   The phenomenological Landau-Ginzburg model introduced by I. Dzyaloshinskii \cite{Dzyaloshinskii} was microscopically derived by T. Moriya \cite{Moriya}.
The DM interaction has been shown to generate skyrmions in thin films \cite{ABR2017,ABR2018a,Diep-Koibuchi} and in magneto-ferroelectric superlattices \cite{Sharafullin1,Sharafullin2,Sharafullin3}.

In this paper, we study a magnetic nanodot embedded in an non-deformable recipient, using the mobile Heisenberg spins. The mobile Potts model with short-range interaction has been used to study the phase transition and the sublimation in a solid \cite{ABR-HTD-MK}. Here we extend this model to the case of a mobile Heisenberg model which includes a long-range dipolar interaction. The dot can be heated to high temperatures to melt. Experimentally, one can imagine periodic arrays of such a dot embedded on a crystal plane and transport properties of itinerant spins across such a plane can be studied in the presence of dots. Itinerant spins are scattered by magnetic dots and desired transport properties can be obtained by modifying dot arrangements.

The purpose of this work is (i) to investigate  the GS configuration in magnetic nanodots taking into account the short-range exchange interaction and the long-range dipolar interaction, (ii) to study the nature of the ordering and the phase transition at finite temperatures in such nanodots, (iii) melting behavior. The methods we employ in this paper are MC simulations with several techniques.

The paper is organized as follows.
Section \ref{GSM} is devoted to the determination of the GS, while section \ref{FTB} shows MC results of finite-temperature behaviors. Concluding remarks are given in section \ref{Concl}.

\section{Ground state}\label{GSM}
\subsection{Mobile spin model}
\label{sec:Assumptions}

Let us consider a recipient of size $N_L=L\times L\times L_z$ with the cubic lattice where $L$ is the $x$ and $y$ linear dimensions, and $L_z$ the size in the $z$ direction. We consider a  number $N_s$ of mobile Heisenberg spins which is less than the total number of lattice sites $N\times N\times N_z$. Therefore, each lattice site can be occupied by a Heisenberg spin or can be empty.  The concentration of spins in the recipient is therefore $c=N_s/N_L$.

We consider the following Hamiltonian containing the exchange interaction between nearest neighbors (NN) and the dipolar interaction
between spins without cutoff:
\begin{eqnarray}
{\cal H} &=& - \sum_{i,j}^{\text{NN}}J_{ij} \mathbf{S}_i\cdot\mathbf{S}_j
\nonumber \\
   &&- D\sum_{i,j}^{\text{all}}
		\left[			 \frac{3(\mathbf{S}_i\cdot\vec{r}_{ij})(\mathbf{S}_j\cdot\vec{r}_{ij})}{r_{ij}^5} - \frac{\mathbf{S}_i\cdot\mathbf{S}_j}{r_{ij}^3}
		\right],\label{eq:ham}
\end{eqnarray}
where $J_{ij}$ denotes the exchange integral between two NN spins $i$ and $j$, $D$ is the dipolar coupling parameter,  $\mathbf{S}_i$ ($|\mathbf{S}_i|=1 \text{ for all } i$) is the spin at the $i$-th site, and $\mathbf{r}_{ij}$ ($r_{ij}=|\mathbf{r}_{ij}|$) is the position vector connecting the spins at the $i$-th and $j$-th sites. The dipolar energy is calculated from the expression included in the Hamiltonian (\ref{eq:ham}) without any numerical approximations; in particular we do not introduce the cut-off radius, since this has been shown \cite{RochaJAP107,Vedmedenko} to affect quantitatively the calculation results in a sensitive manner.

\subsection{Method of ground-state determination}\label{GS-subs}

In the following, we take $J_{ij}=J_\perp=1$ between NN in the $z$ direction (\emph{i.e.} between NN in adjacent planes) and  $J_{ij}=J_{\para}=4$ between in-plane NN. As will be seen below, another choice will not change our results but it changes the range of values of $D$ to have a vortex GS.

\par
%
%
\par
We start from a random spin configuration of $N_s$ mobile spins. By random, we mean spins occupy random lattice positions and have random orientations. This state corresponds to a high-$T$ disordered phase. We slowly cool the system from that state using the Metropolis algorithm \cite{Landau09,Brooks11}.  The spins move from site to site, and at low $T$, they condense into a film with a number of successive layers fully occupied. This is due to the fact that the in-plane interaction is much larger than the perpendicular interaction. However, due to the long-range nature of dipolar interaction, the spin orientations are still in a weak disorder showing no symmetric configuration expected from the symmetry of the Hamiltonian.
To get rid of this and to come down to $T=0$, we need to minimize locally the energy of each spin as follows.
We consider here a spin localized on the lattice site at $T=0$. To find the ground state (GS) of the system we minimize the energy of each spin, one after another. This can be numerically achieved as the following. At each spin, we calculate the local-field components acting on it from its NN using the above equations. Next we align the spin in its local field, \emph{i.e.} taking $S_i^x=H_i^x/\sqrt{\left(H_i^x\right)^2+\left(H_i^y\right)^2+\left(H_i^z\right)^2}$ etc. The denominator is the modulus of the local field. In doing so, the spin modulus is normalized to be 1. As seen from Eq. (\ref{eq:ham}), the energy of the spin $\mathbf S_i$ is minimum. We take another spin and repeat the same procedure until all spins are visited. This achieves one iteration. We have to do a sufficient number of iterations until the system energy converges.  The spins at $T=0$ form a dot of size $L\times L\times L_s$ which verifies $c=N_s/N_L=L_s/L_z$.

\begin{figure}[h!]
\center
\includegraphics[width=5.5cm]{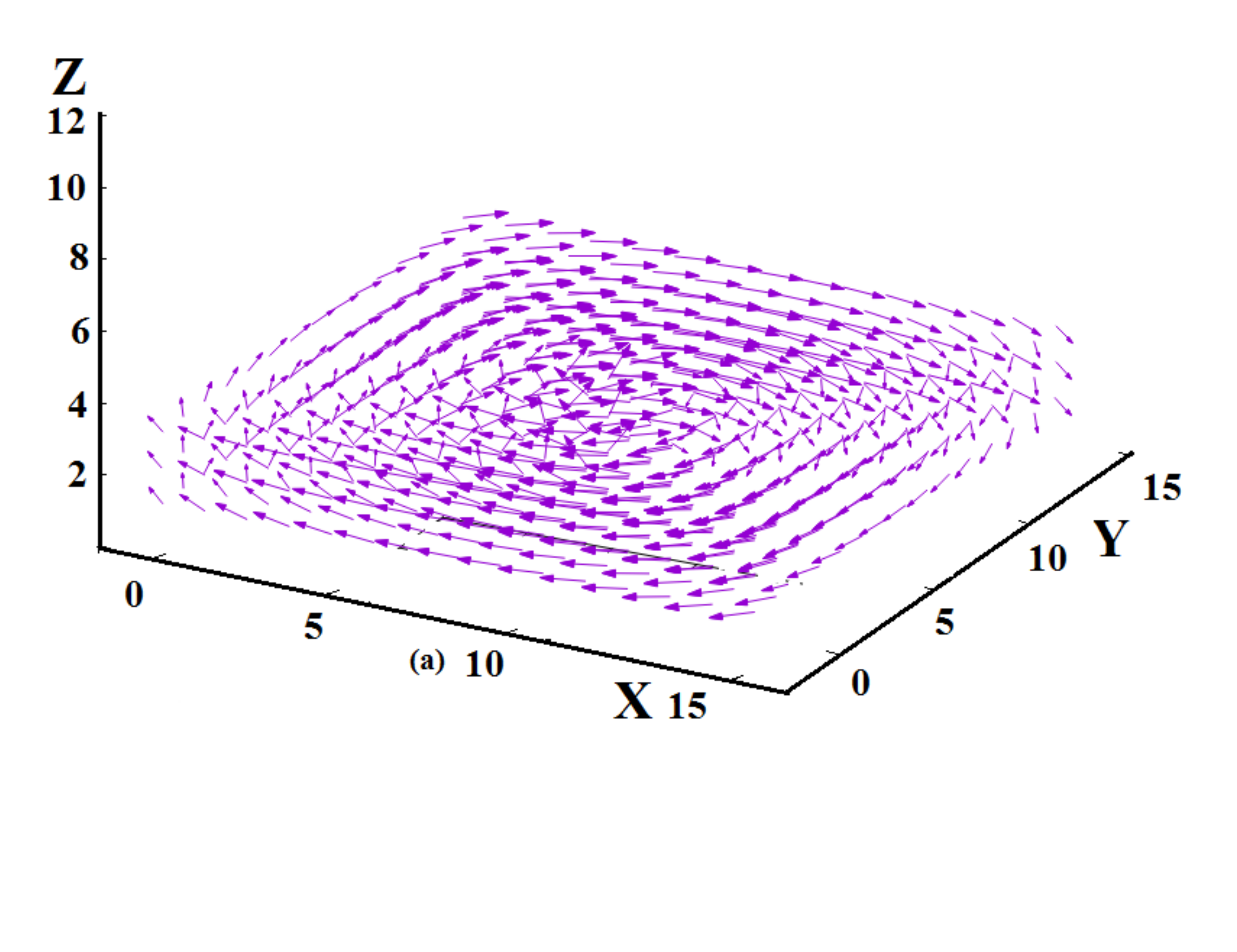}
\includegraphics[width=5.5cm]{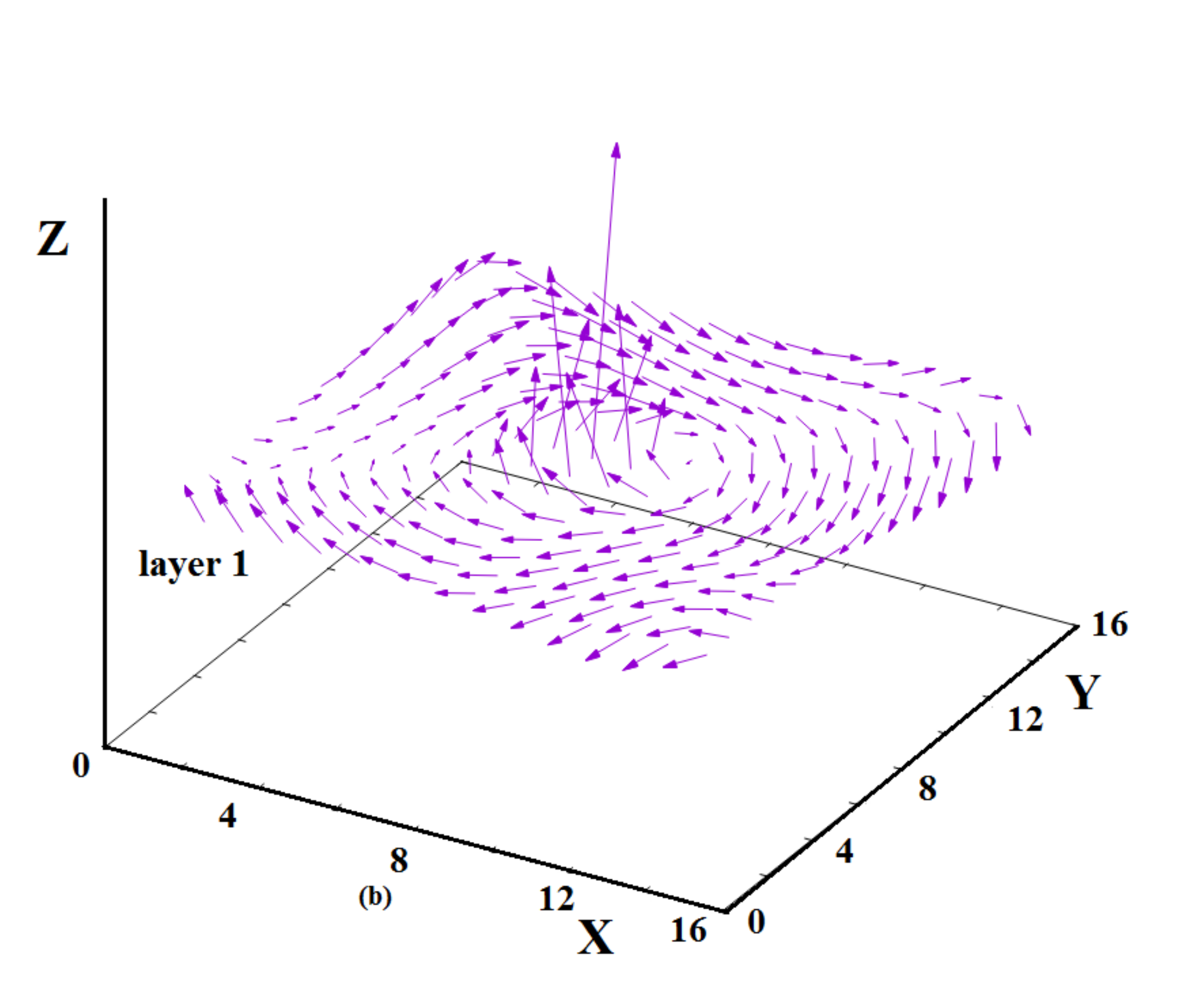}
\includegraphics[width=5.5cm]{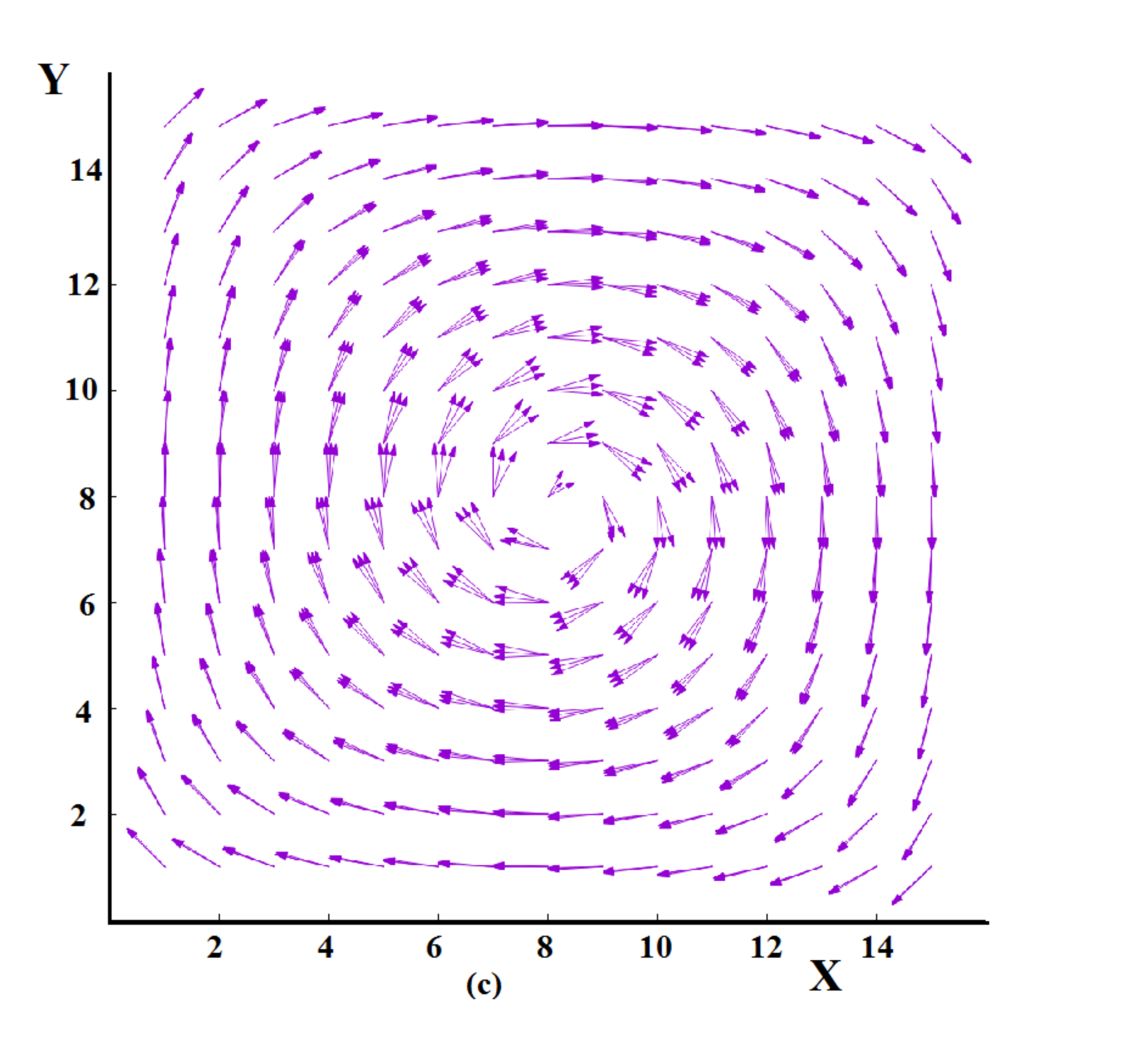}
\includegraphics[width=5.5cm]{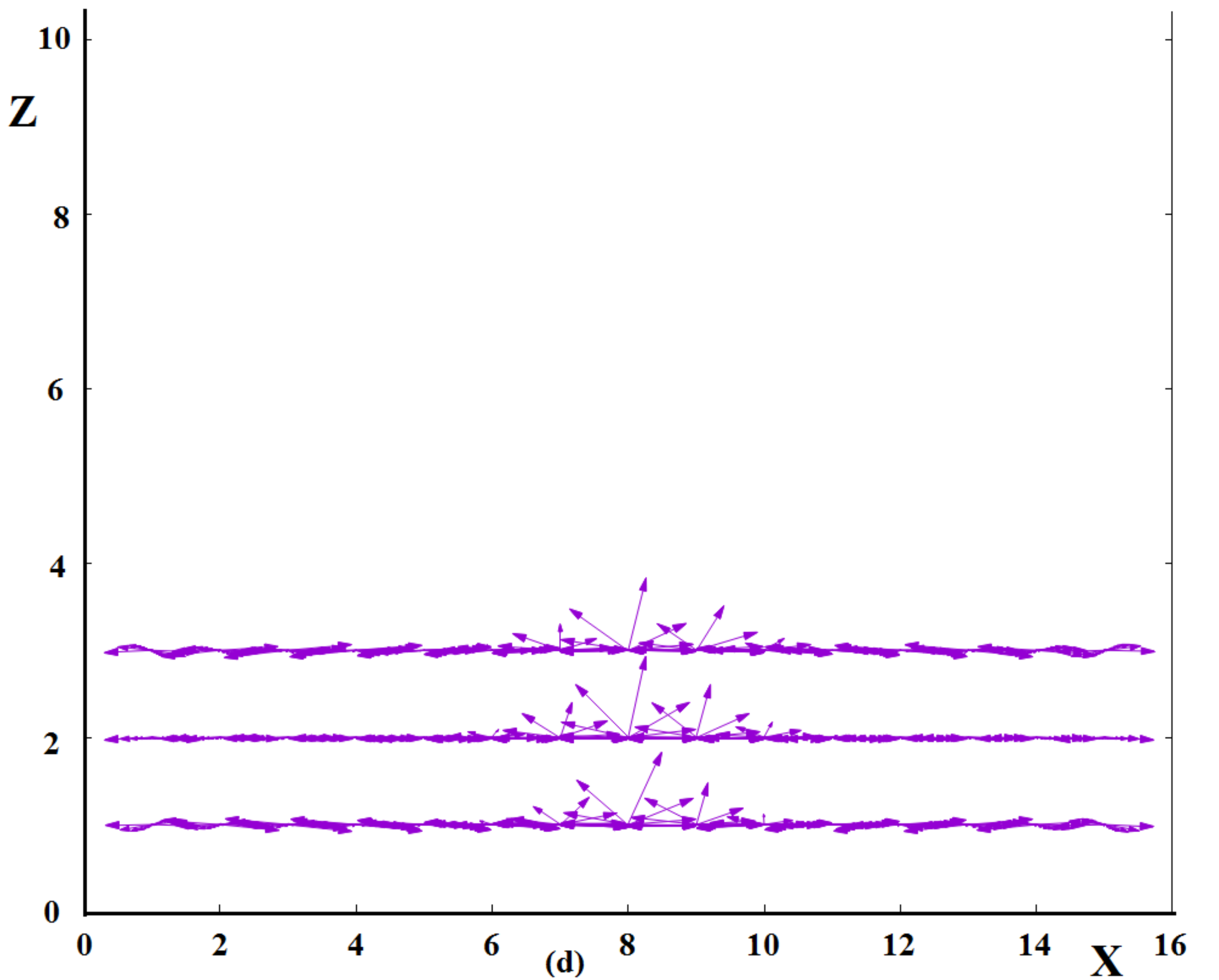}
\caption{Ground state of $c=25\%$ of mobile spins in a recipient of $15\times 15\times 12$. The dot at $T=0$ has the size $15\times 15\times 3$. We have used  $J_\perp=1$, $J_{\para}=4$, $D=1$: (a) 3D view, (b) First-layer configuration, (c) Projection on the $xy$ plane, (d) Projection on the $xz$ plane. See text for comments. \label{GS1}}
\end{figure}

An example of  GS are displayed in Fig. \ref{GS1} in a recipient of $15\times 15\times 12$. The concentration used is $c=25\%$.  We see that the dot size at $T=0$ is $15\times 15\times 3$.
Let us give some comments on Fig. \ref{GS1}. In Fig. \ref{GS1}a the dot is viewed in 3D space. We see three compact layers. Figure \ref{GS1}b shows the first layer structure, Fig. \ref{GS1}c shows the projection on the $xy$ plane where we can see that the projection of the spins near the dot center are not exactly parallel. Figure \ref{GS1}d shows the projection on the $xz$ plane, namely a side view of the dot. One sees that the spins at the dot center go out of the $xy$ plane. This is easily understood  in order these spins reduce the spin orientation constraint. In a zero field in the $z$ direction, these central spins can point in the + or -$z$ direction. There is therefore a two-fold degeneracy. It may have application in magnetic recording memory using a small magnetic field along the $\pm z$ direction to control the magnetization direction.

\begin{figure}[h!]
\center
\includegraphics[width=4cm]{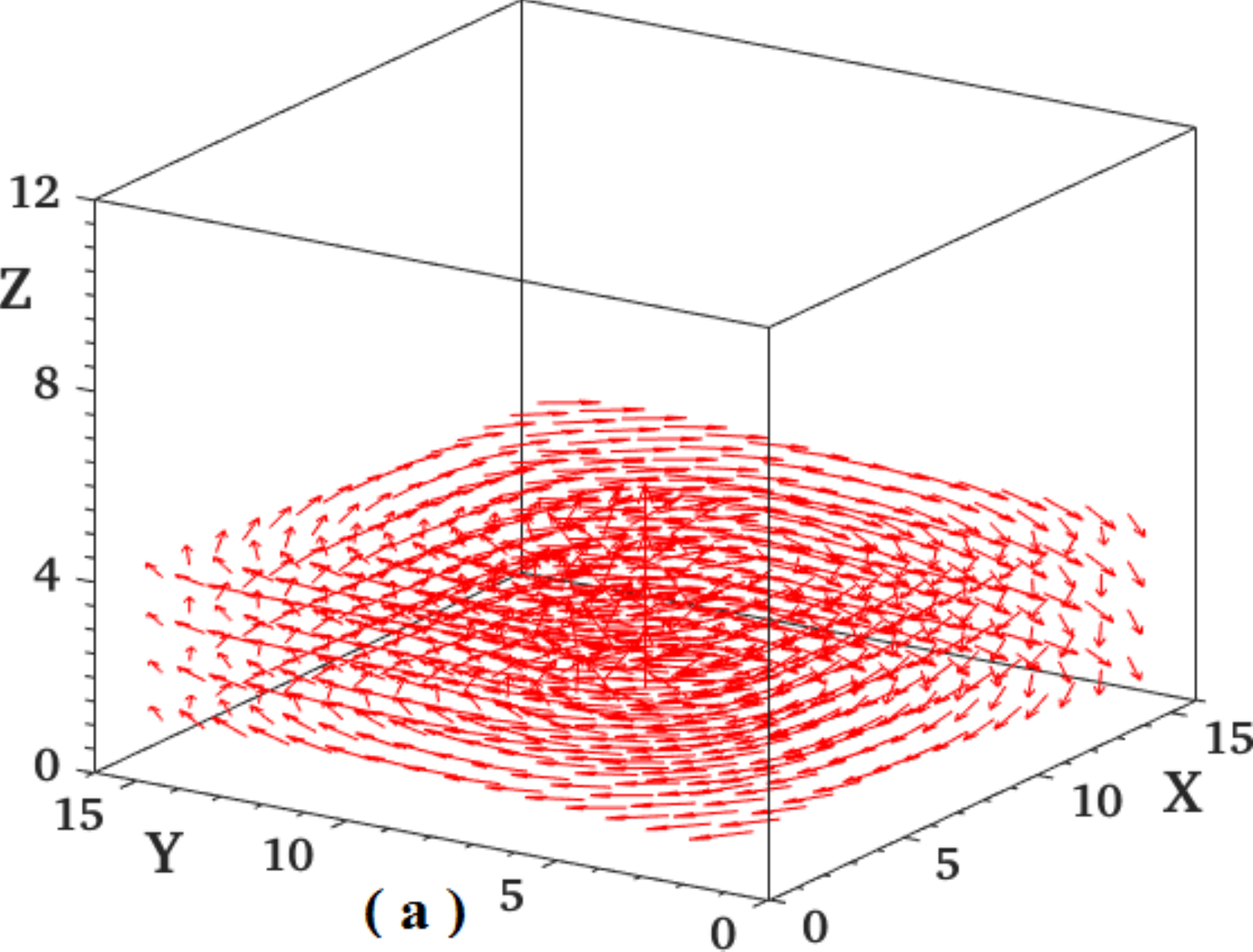}
\includegraphics[width=3.5cm]{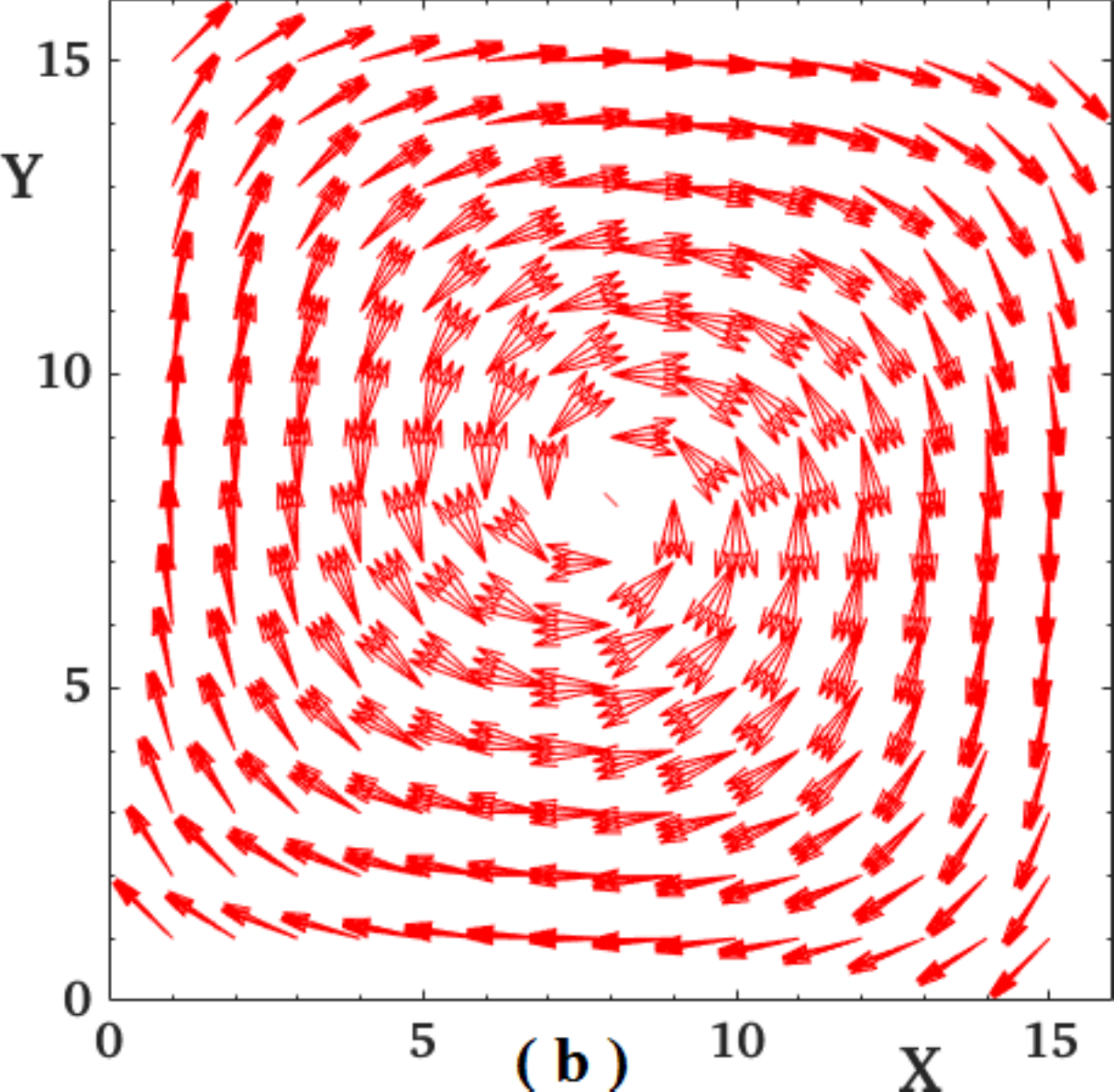}
\includegraphics[width=4cm]{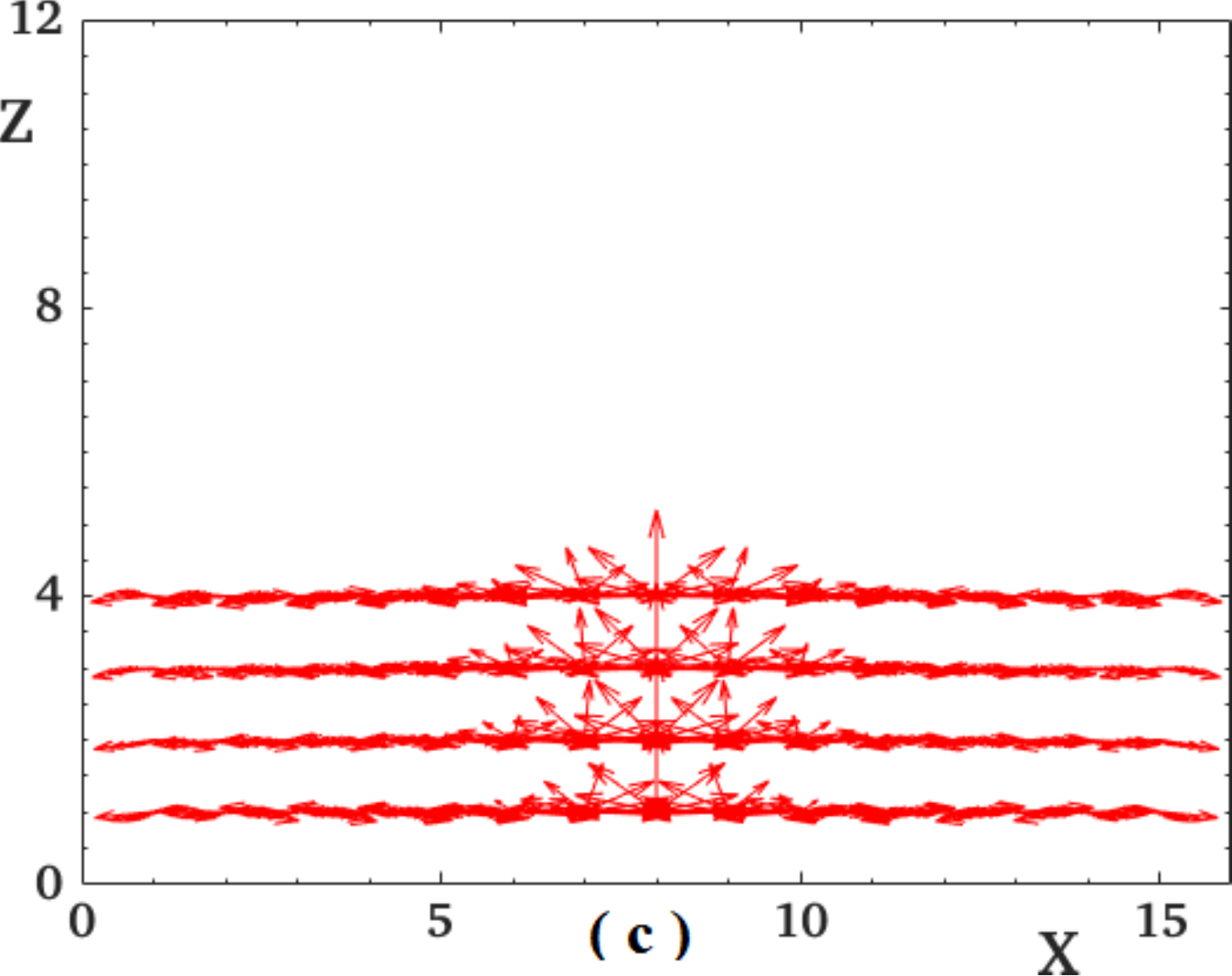}
\caption{Ground state of $c=30\%$ of mobile spins in a recipient of $15\times 15\times 12$. The dot at $T=0$ has the size $15\times 15\times 4$ (four layers). We have used  $J_\perp=1$, $J_{\para}=4$, and with a dipolar interaction $D = 0.7$: (a) 3D view, (b) Projection on the $xy$ plane, (c) Projection on the $xz$ plane to see the spin $z$-components.
The GS exhibits a vortex
around the center of the dot. The spins lie in the $xy$ plane at the
border except around the vicinity of the center where they have
non-zero $z$ components.\label{GS2}}
\end{figure}

Note that there is a range of $D$ where we observe such vortex structure for a given ($J_\perp,J_{\para}$). When we change the recipient size and the concentration, this range of $D$ changes. What we mean by this sentence is that when we increase the system size for example, in the dipolar term a spin interacts with a larger number of spins because there is no cutoff in the dipolar sum of Hamiltonian (\ref{eq:ham}).
Therefore, to obtain a vortex which is favored by the dipolar interaction, we just need a smaller $D$ to overcome the energy of the linear configuration favored by a given set of ($J_\perp, J_{\para}$). Another example of concentration is shown in Fig. \ref{GS2}.
Other cases displayed in Figs. 3-6 below show that when we decrease the concentration $c$ we have to increase $D$ and vice-versa in order to have a vortex structure.  It is not our purpose to precisely determine the range of $D$ in which there is a vortex configuration. We give here only the physical mechanism which determines the planar vortex phase.

\section{Melting}\label{FTB}

\subsection{Slow heating}
In this section, we slowly heat the dot in a recipient. We take the recipient height much larger than the thickness of the dot. The empty space in the $z$ direction allows the dot to melt into a liquid at high temperature ($T$).

Since the spins are mobile, a spin can move from one lattice site to a nearest site.  MC simulation is used to update the spin position and the spin orientation. The spin position is updated whenever there are vacant sites next to it, at the same time with the orientation update, using the Metropolis algorithm.

We start the simulation using the GS state configuration and we heat the system to a temperature $T$.  As usual, we discard a large number of MC steps to equilibrate the system before averaging physical quantities over a large number of MC steps. Physical quantities which are calculated include
 the energy $E$, the magnetization, the heat capacity and the magnetic susceptibility
as functions of $T$ for different concentrations $c$, different sizes of the recipient. We have also calculated
 the diffusion coefficient, which is the sum of the mean square of the distance made by each spin at each $T$.
We have also computed the mean value of the number of nearest neighbors as a function of $T$.  We recall some definitions:

The total energy \\
\begin{equation}
 {E} = \langle \mathcal{H} \rangle \\
\end{equation}
The Edwards-Anderson order parameter $Q_{EA}$ is \\
\begin{equation}
Q_{EA} = \frac{1}{N_s (t_a - t_0)} \sum_{i}\vert \sum_{t=t_0}^{t_a} \mathbf{S}_i(t)\vert\\
\end{equation}
where $\langle \cdot \rangle$ indicates the thermal average with $N_{s}$ being the total number of spins. Note that the
$Q_{EA}$ is calculated by taking the time average of each spin before averaging over all spins of the system.  This order parameter is very useful in the case of disordered systems such as spin glasses \cite{Mezard} or doped compounds: it expresses the degree of freezing of spins independent of whether the system has a long-range order or not \cite{DiepTM}.

To study the change of behavior, we will show the results of the following situations for comparison:

i) all spins are supposed to be localized on their site (this case corresponds to the solid crystal)

ii) spins on one, two, three, ... layers are mobile (this case allows us to study the partial melting process)

iii) all spins are mobile. This case corresponds to the full melting to the liquid phase at high enough $T$.

\subsection{Concentration effects}\label{con-eff}

As said in the previous section, when we change the concentration, we need another value of $D$ to obtain a vortex configuration. For 3 layers ($c=25\%$), we pick $D=1$ in the range of the vortex phase, while for 4 layers ($c=30\%$), we pick $D=0.7$ which lies in the middle of the vortex phase. This explains why we have to specify $D$ for each concentration. Varying $D$ slightly around the chosen values does not change qualitatively the results.
Now if we work with one fixed variable, say $c$, and we study the system behavior as a function of $D$, we find the vortex phase in a range of $D$, the collinear configurations at small $D$, and no planar vortex at very high $D$. The linear phase and the non-planar vortex phase are not interesting with respect to applications using the spin reversal by a small magnetic field.

\begin{figure}[h!]
\centering
\includegraphics[scale=0.40]{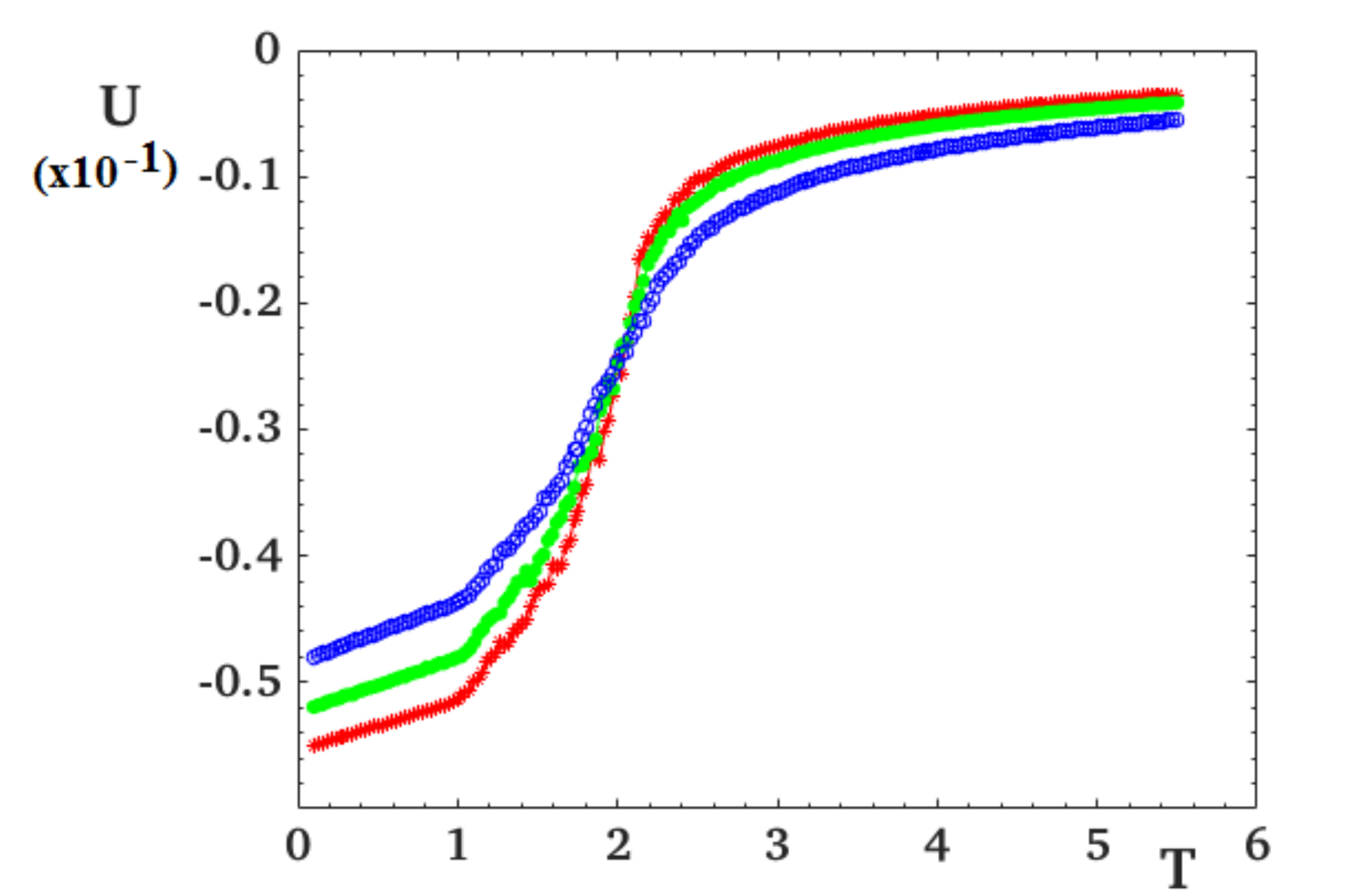}
\caption{Energy per spin $U$ vs $T$. Effect of the concentration. The lattice size of the system is $15 \times 15 \times 12$ with $J_{\para}=4$ and $J_\perp = 1$. The blue curve corresponds to a spin concentration equal to $c=50\%$ with a magnetic dipole-dipole interaction equal to $D=0.3$. The green curve corresponds to $c=30\%$ and $D=0.7$. The red curve corresponds to $c=25$\% and $D=1$.\label{E1}}
\end{figure}

Let us show in Fig.\ref{E1} the results of the energy as a function of $T$ for various dot thickness, namely various concentrations $c$ with the choice of $J_{\para}=4$ and $J_\perp = 1$. All spins are mobile.

\begin{figure}[h!]
\centering
\includegraphics[scale=0.40]{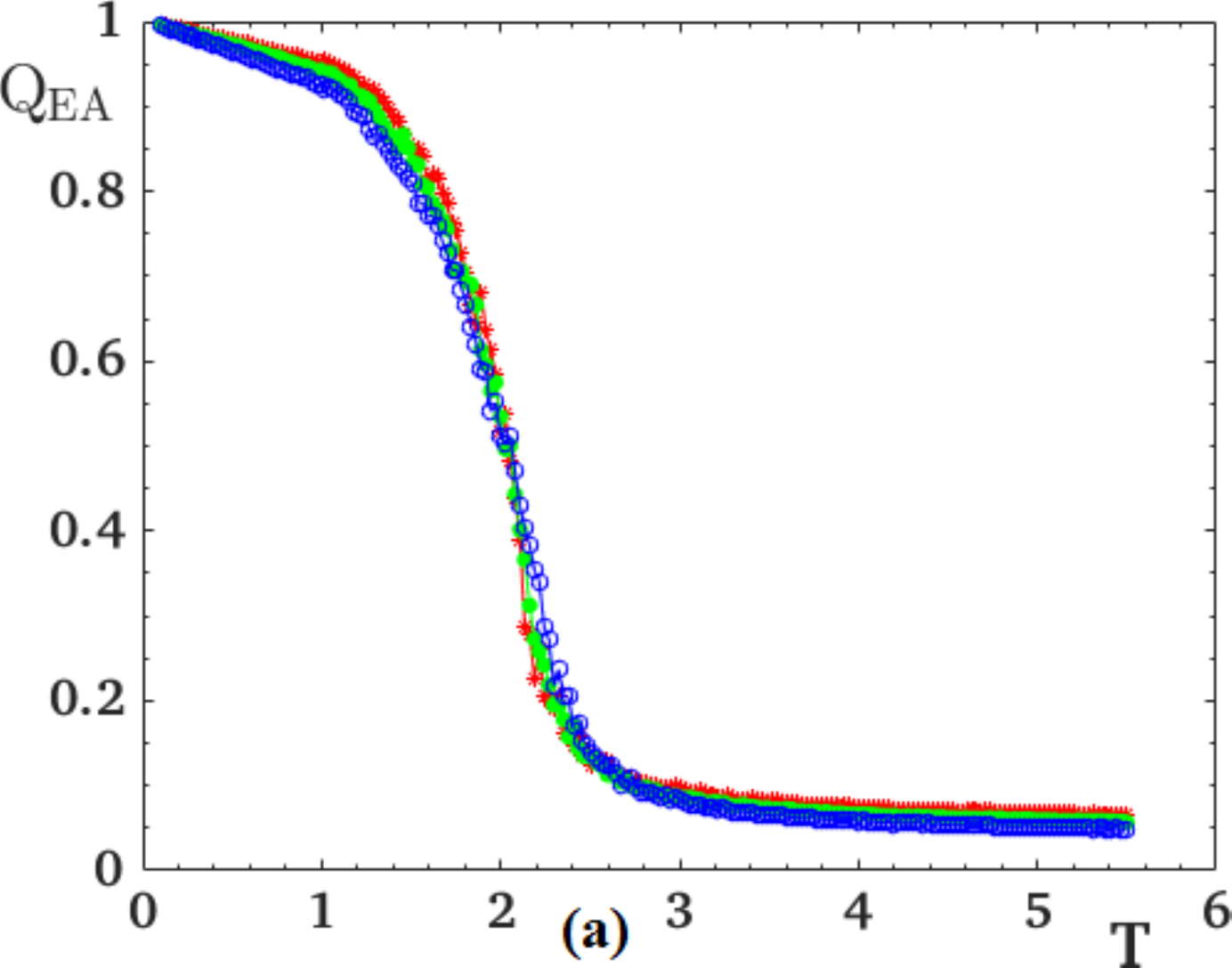}
\includegraphics[scale=0.40]{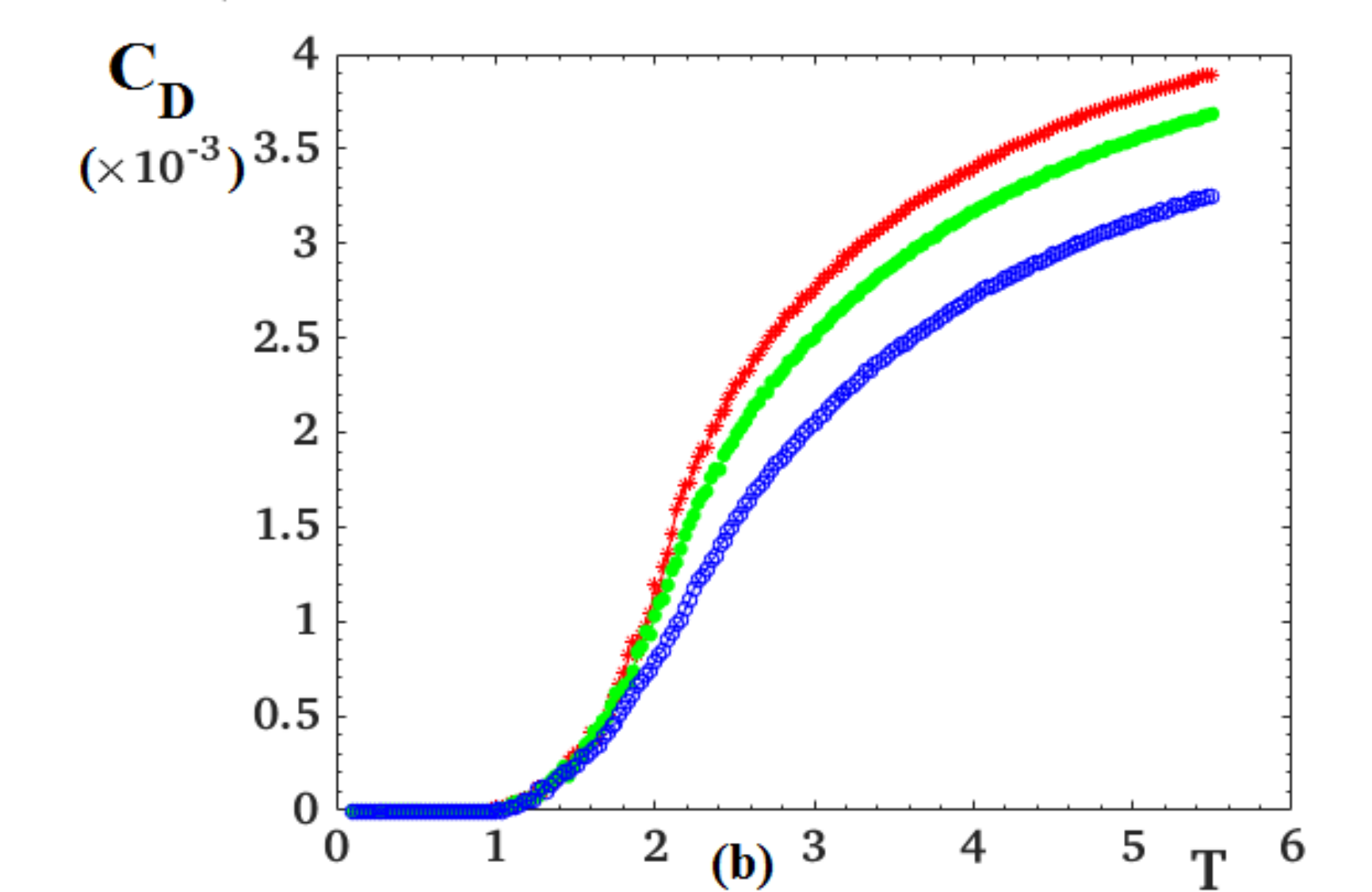}
\caption{(a) The Edwards-Anderson order parameter $Q_{EA}$ vs $T$ for several concentrations, (b) The diffusion coefficient $C_D$ vs $T$.  The lattice size of the system is $15 \times 15 \times 12$ with $J_{\para}=4$ and $J_\perp = 1$. The blue curve corresponds to $c=50\%$ and $D=0.3$. The green curve corresponds to $c=30\%$ and $D=0.7$. The red curve corresponds to $c=25$\% and $D=1$.}\label{EAD}
\end{figure}

It is interesting to note that the three energy curves have the same transition temperature $T_c$, a kind of fixed point independent of the concentration $c$.  The Edwards-Anderson order parameter is shown in Fig. \ref{EAD}a confirms that the transitions of these concentrations occur at the same temperature within statistical errors.  We show in Fig. \ref{EAD}b the diffusion coefficient. We see here that spins start to move after the transition at $T_c$ from the vortex configuration to the disordered phase.

The fact that the meting temperature does not change with the concentration up to 50\% means that spins have enough empty space to evaporate. We may imagine the extreme situation where $c$ is close to 1: the first evaporated spins in the little empty space prevent the following spins to evaporate due to the lack of empty sites. This increases the melting temperature.

\subsection{Comparison with a localized-spin system}

\begin{figure}[h!]
\centering
\includegraphics[scale=0.40]{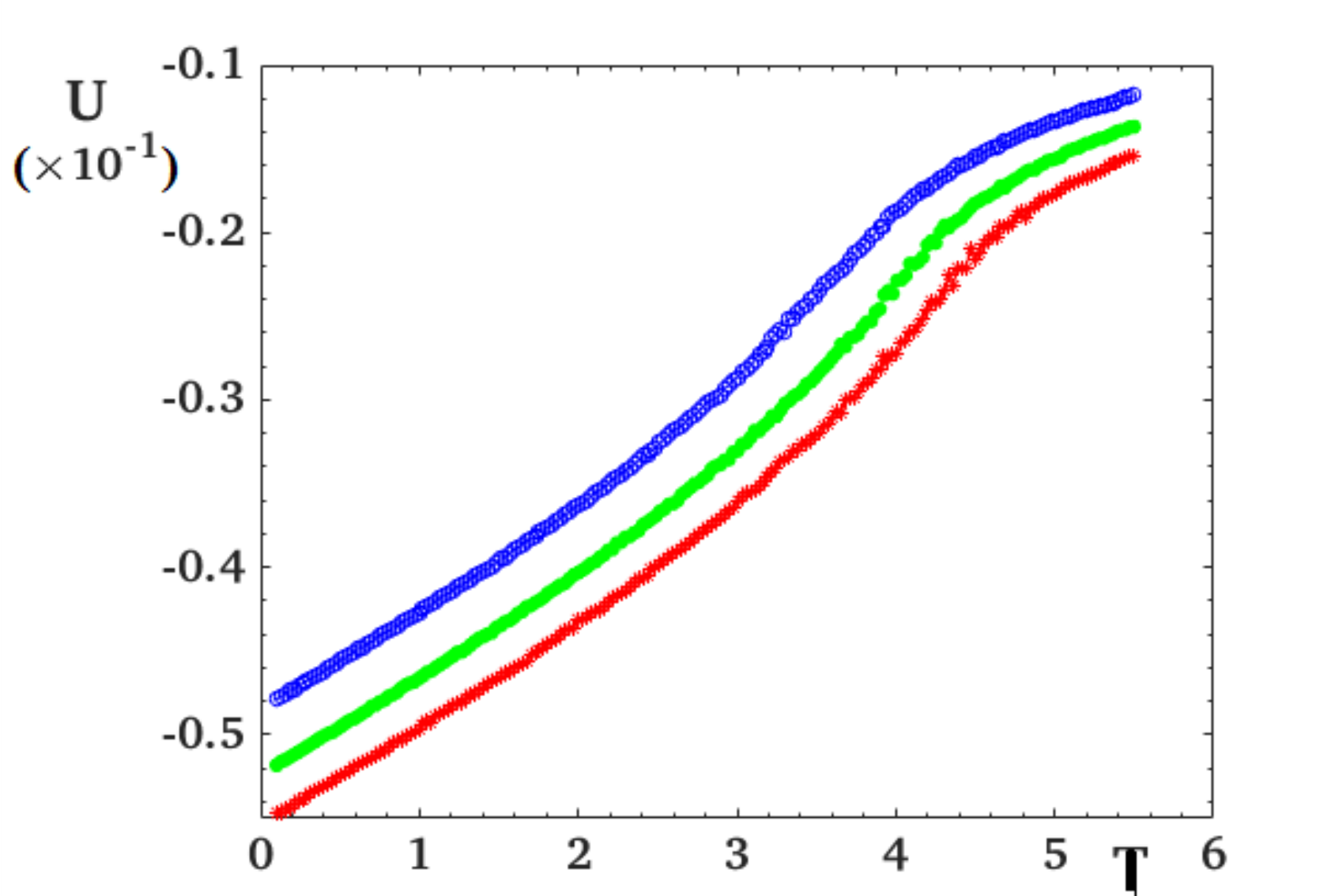}
\caption{Energy vs $T$. Localized spin model. The lattice size of the system is $15 \times 15 \times 12$ with  $J_{\para}=4$ and $J_\perp = 1$. The blue curve corresponds to $c=50\%$, \emph{i.e.} a 6-layer film, with a magnetic dipole-dipole interaction equal to $D=0.3$. The green curve corresponds to $c=30\%$ (4-layer film) and $D=0.7$. The red curve corresponds to $c=25$\% (3-layer film) and $D=1$.}\label{E2}
\end{figure}

We show here the case of a system of localized spins to compare with the mobile spins shown above.  Figure \ref{E2} shows the energy for three thicknesses  of the dot.  Unlike the full melting case shown above, the transition occurs at a different temperature for a different thickness. This is well-known in magnetism of thin solid films: the transition temperature increases with increasing thickness \cite{DiepTM,Pham2009}.  We come back to this point at the end of this subsection.

In addition to the Edwards-Anderson order parameter, we can define an order parameter in the case of no long-range GS ordering: if the GS is well defined by a numerical method, then we can project the actual spin configuration of at a given $T$ at a given time $t$ on the GS. Needless to say, if the spin configuration is not strongly deviated from the GS, the order parameter is close to 1. This is defines as:

\begin{equation}\label{OP}
P(T)=\frac{1}{N_s (t_a-t_0)}\sum_i |\sum_{t=t_0}^{t_a}  \mathbf{S}_i (T,t)\cdot \mathbf{S}_i^0(T=0)|
\end{equation}
where $\mathbf{S}_i (T,t)$ is the $i$-th spin at the time $t$, at temperature $T$, and $\mathbf{S}_i^0 (T=0)$ is its state in the GS. The order parameter $P(T)$ is close to 1 at very low $T$ where each spin is only weakly deviated from its state in the GS. $P(T)$ is zero when every spin strongly fluctuates in the paramagnetic state.
The above definition of $P(T)$ is similar to the Edward-Anderson order parameter used to measure the degree of freezing in spin glasses \cite{Mezard}: we follow each spin with time evolving and take the spatial average at the end.  However, the advantage of $P(T)$ is the fact that we can follow the GS configuration until it is broken.

\begin{figure}[h!]
\centering
\includegraphics[scale=0.40]{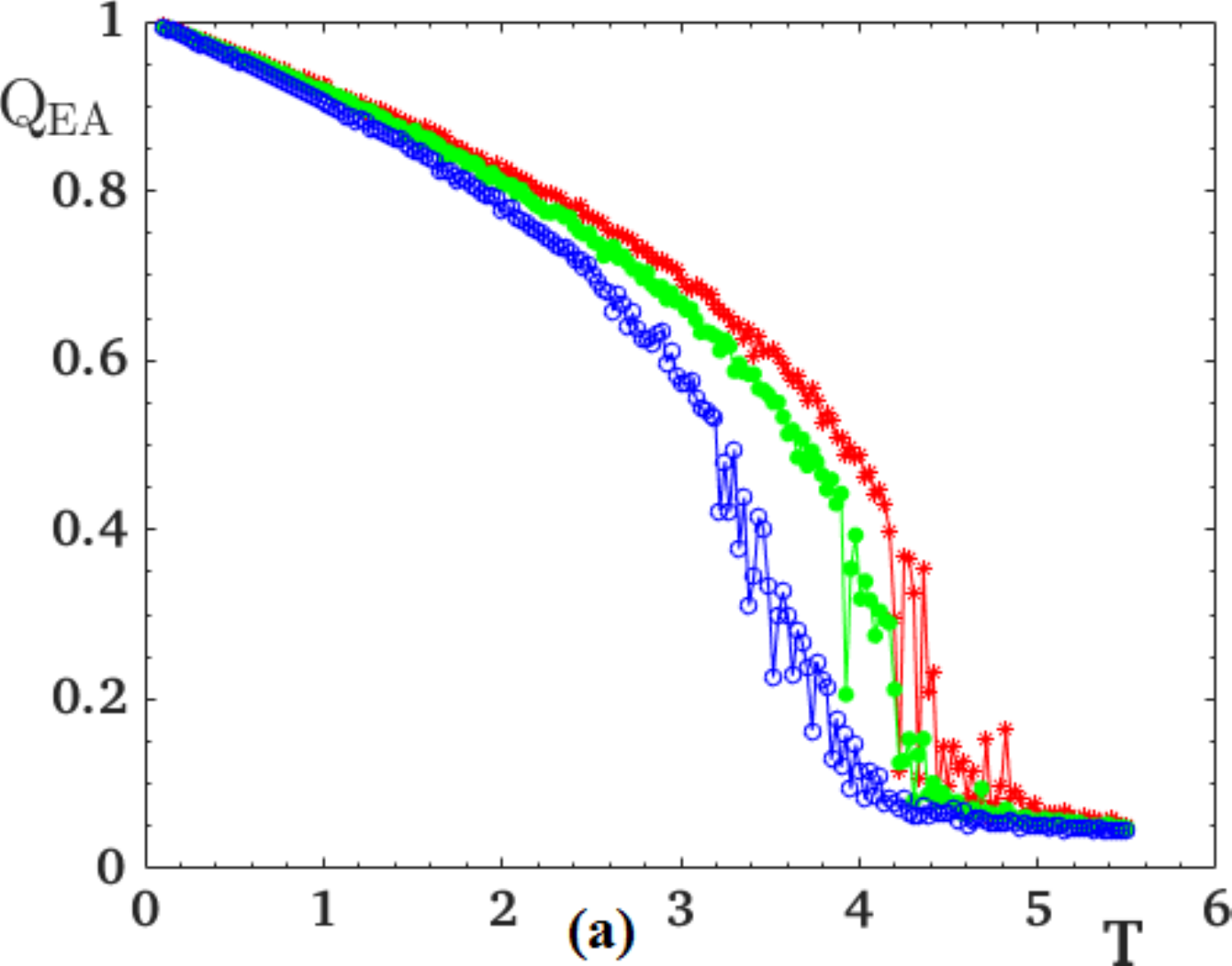}\\
\includegraphics[scale=0.40]{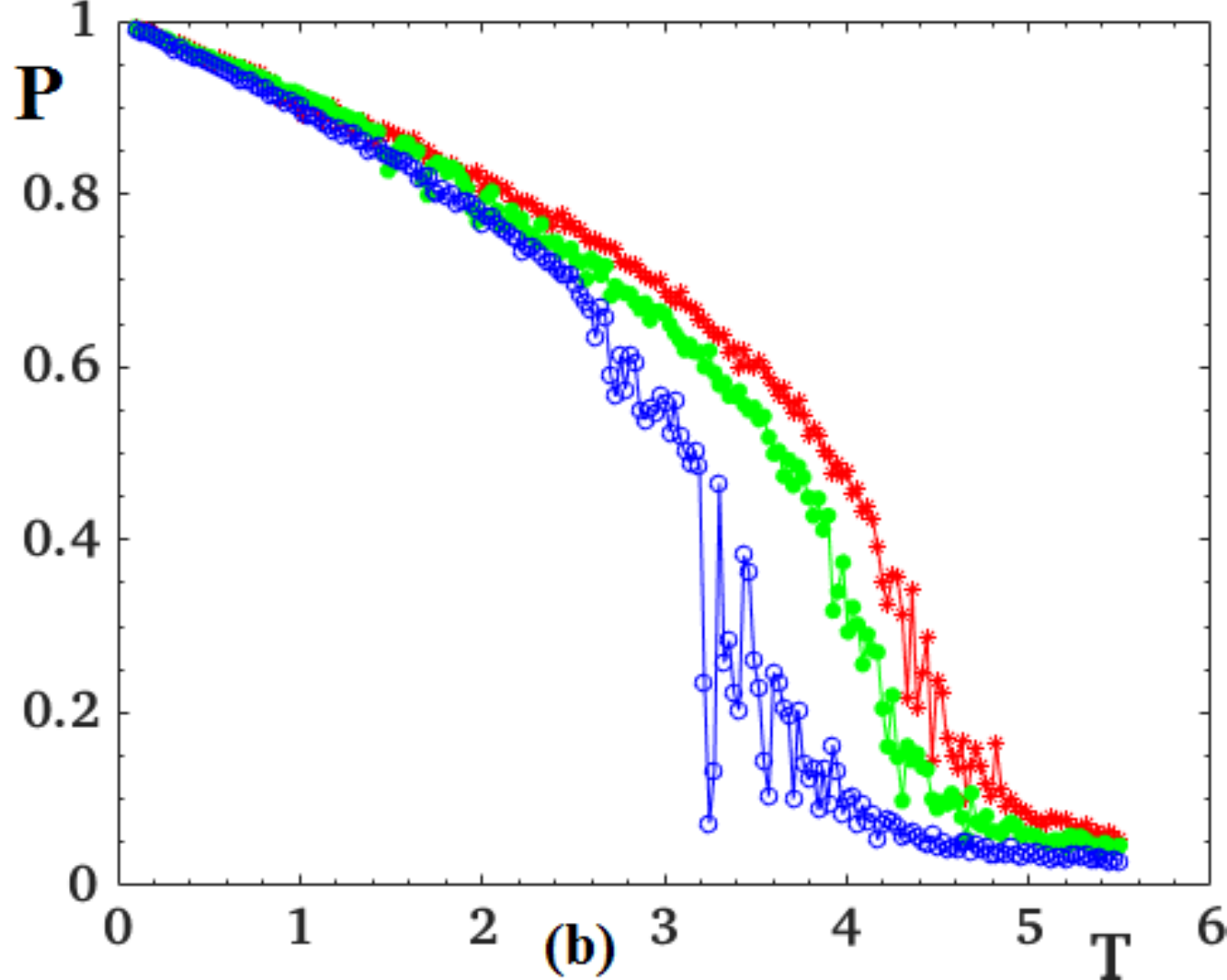}\\
\caption{(a) Edwards-Anderson order parameter  vs $T$, (b) Order parameter $P$. The model is the localized spin model. The lattice size of the system is $15 \times 15 \times 12$ with  $J_{\para}=4$ and $J_\perp = 1$. The blue curve corresponds to a 6-layer film, with a magnetic dipole-dipole interaction equal to $D=0.3$. The green curve corresponds to a 4-layer film and $D=0.7$. The red curve corresponds to a 3-layer film and $D=1$.}\label{EAD2}
\end{figure}

We show in Fig. \ref{EAD2} the Edwards-Anderson order parameter $Q_{EA}$ and the projection order parameter $P$ as functions of $T$.  These confirm the difference of $T_c$ for different film thicknesses. Note however that in magnetic thin films with localized spins, $T_c$ increases with increasing thickness provided that all parameters other than the film thickness are the same \cite{Pham2009}. In the present model, we cannot have this condition because we should choose $D$ to produce the vortex structure while varying the film thickness, explained previously in sections \ref{GS-subs} and \ref{con-eff}. As a matter of fact,  Figs. \ref{E2} and \ref{EAD2} show the data for three different thicknesses with three different $D$, i. e. for three different systems. The thickness is not the only variable here. So the variation of $T_c$ is not due to it alone, the scaling in Ref. \cite{Pham2009} does not apply.

\subsection{Effect on the number of layers that can melt}

Let us show now the case where spins in one, two or more layers are mobile.  This allows us to follow the melting progressively. Though artificial, this procedure corresponds to a reality when the surface layer melts first, then the second layer, ... with increasing $T$ \cite{Bocchetti}.

\begin{figure}[h!]
\centering
\includegraphics[scale=0.40]{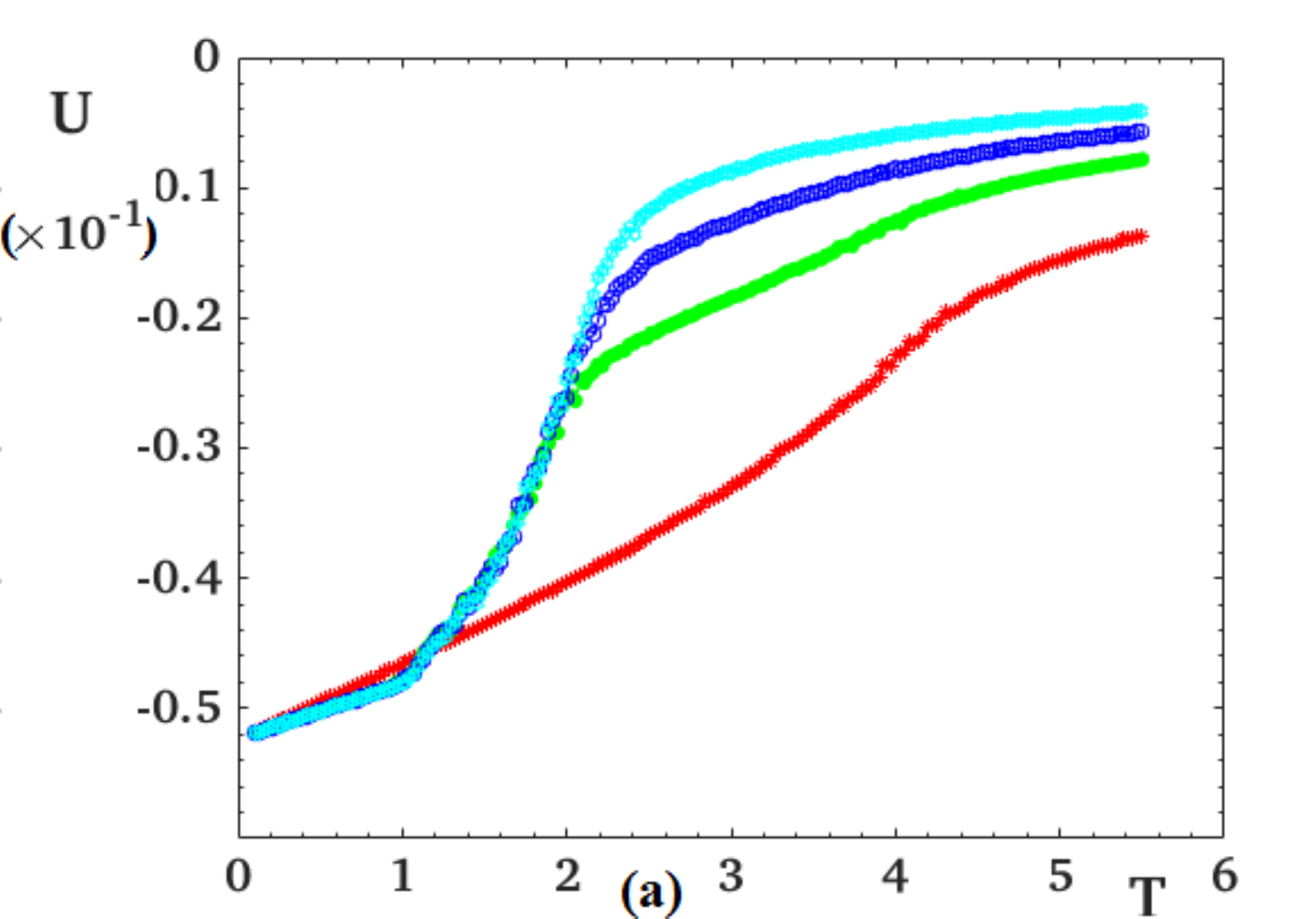}
\includegraphics[scale=0.40]{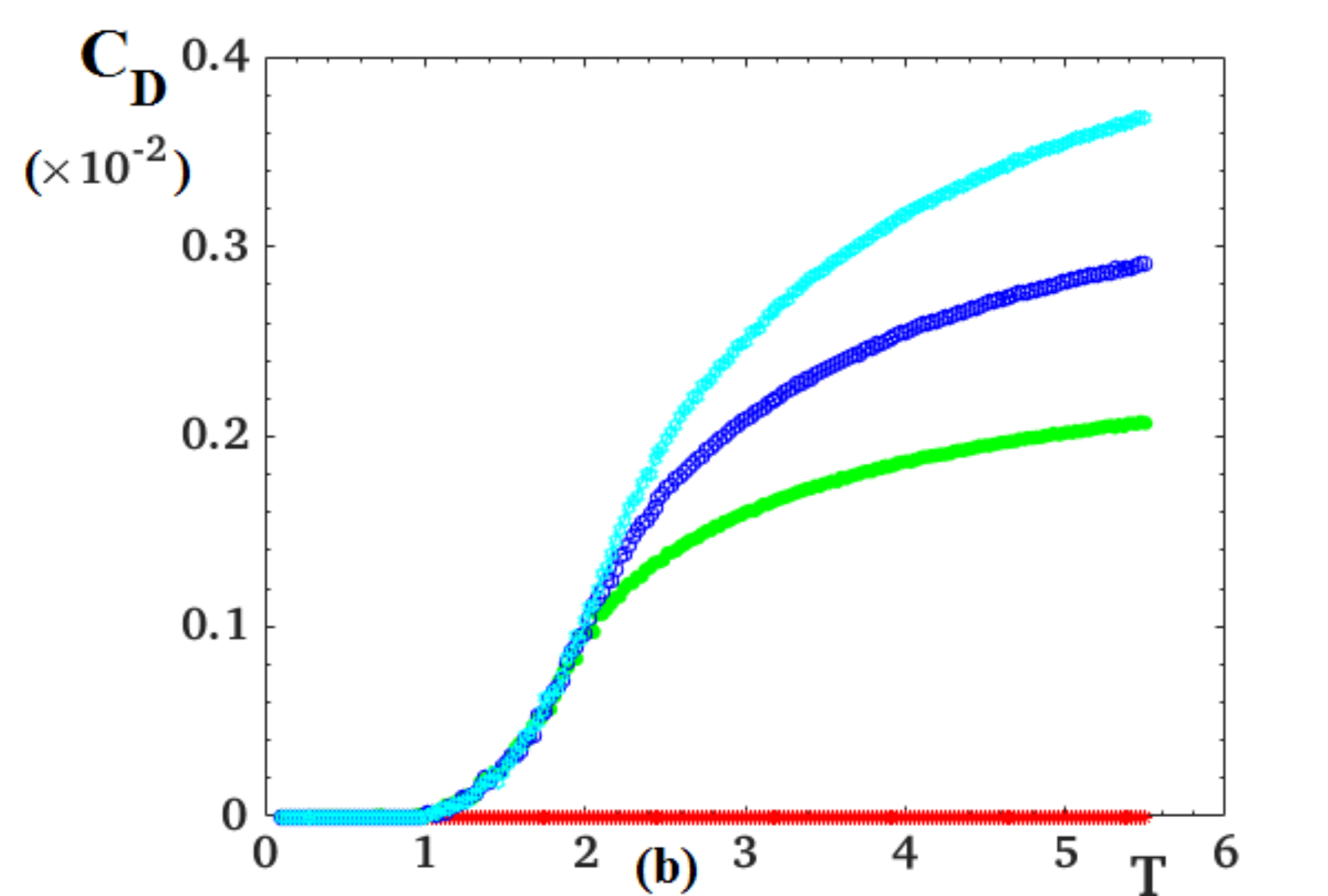}
\caption{(a) Energy $U$ vs $T$, (b) Diffusion coefficient $C_D$ vs $T$.  The lattice size of the system is $15 \times 15 \times 12$ with $J_{\para}=4$, $J_\perp = 1$ and $D=0.7$. The concentration is fixed to $c=30\%$, \emph{i.e.} 4 filled layers. The red curve  corresponds to a localized-spin system. The green curve corresponds to the two first layers can melt; the blue curve to three mobile layers and the cyan curve to four mobile layers.}\label{E3}
\end{figure}

Figure \ref{E3} shows the energy and the diffusion coefficient versus $T$.
The case of localized spins is also shown for comparison.  One sees that the transition
temperature decreases as the number of mobile layers increases. Note that the case of a system of completely localized spins has a very high $T_c$.

\begin{figure}[h!]
\centering
\includegraphics[scale=0.40]{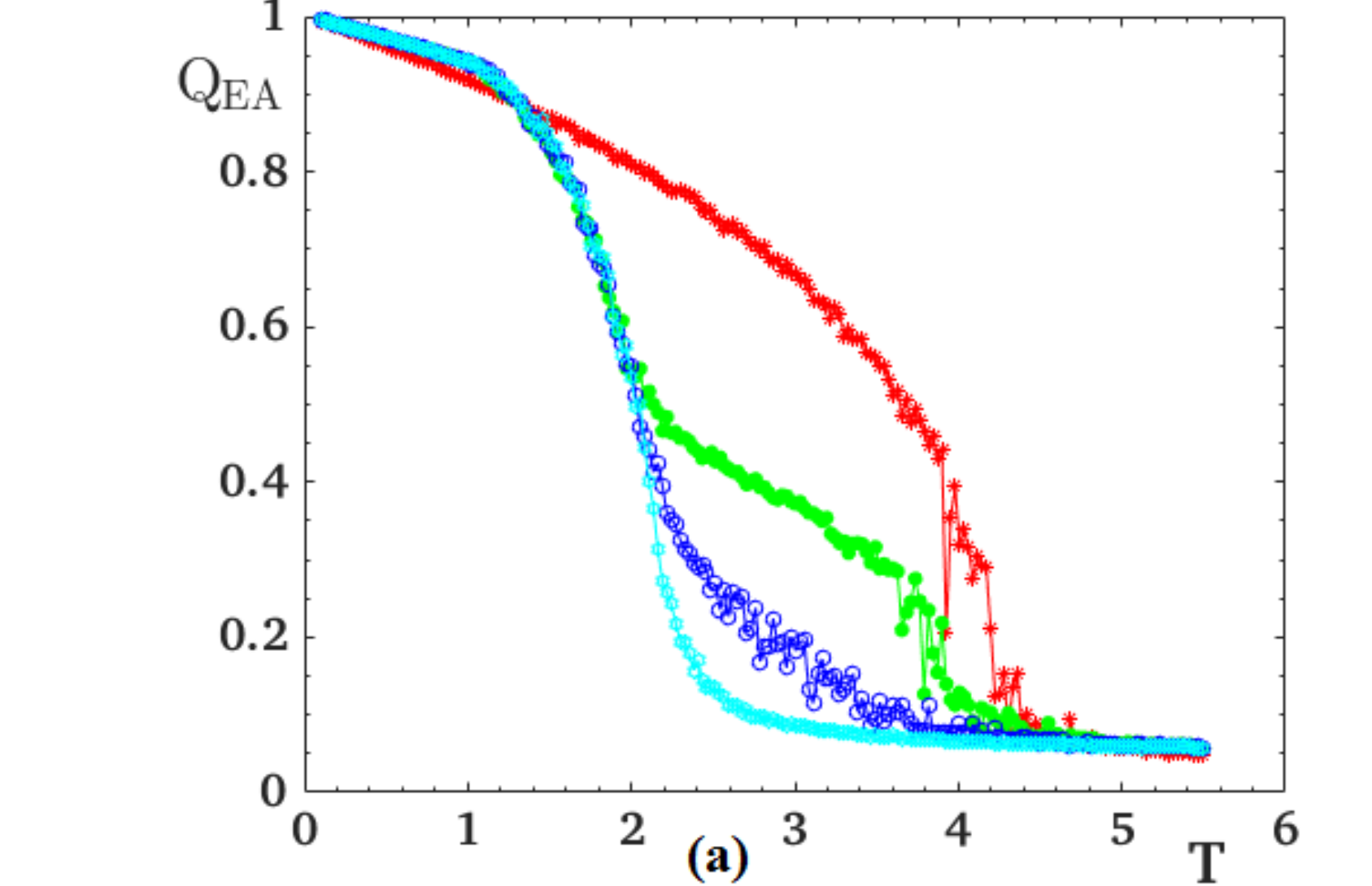}\\
\includegraphics[scale=0.40]{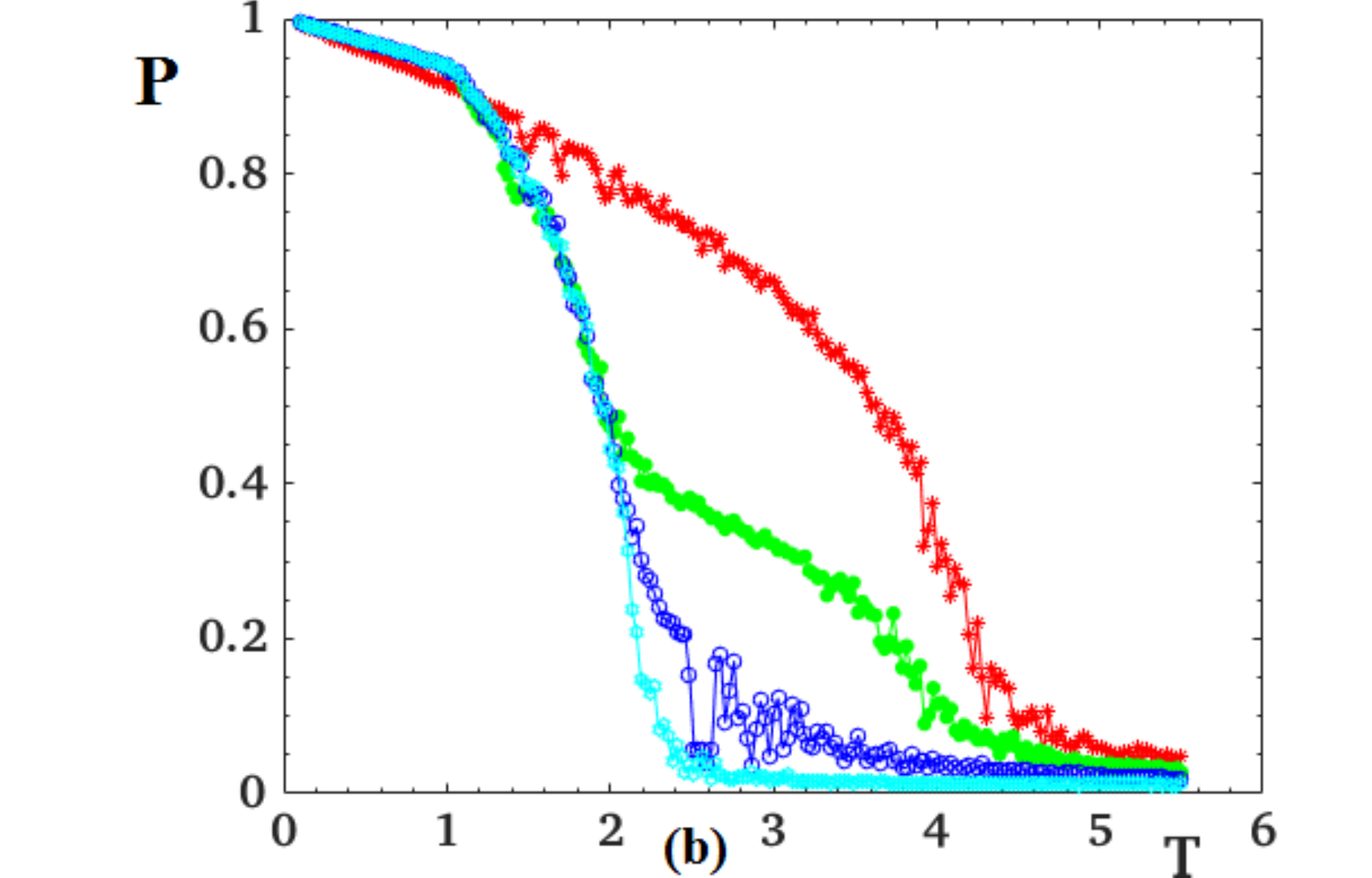}\\
\caption{System with partially mobile layers. (a) Edwards-Anderson order parameter vs $T$, (b) Projection of the order parameter $P$ vs $T$.  The lattice size of the system is $15 \times 15 \times 12$ with  $J_{\para}=4$, $J_\perp = 1$, and $D=0.7$. The concentration is fixed to $c=30\%$, \emph{i.e.} 4 filled layers. The red curve  corresponds to a localized-spin system, the green curve corresponds to the case where the two first layers (starting from the top) are constituted with mobile spins, the blue curve corresponds  to the case of three mobile layers,  and the cyan curve to four mobile layers.}\label{EAD3}
\end{figure}

Figure \ref{EAD3} shows the Edwards-Anderson order parameter and the order parameter $P$.
Several remarks are in order:
(i) for a system of completely localized spins or a system of completely mobile spins, there is only one transition for each concentration, (ii) when a number of layers are mobile, the system has a partially ordered state: the mobile layers become disordered at some $T$ but the localized layers are still ordered. This causes a step in the order parameters observed in Fig. \ref{EAD3}.

\begin{figure}[h!]
\centering
\includegraphics[scale=0.40]{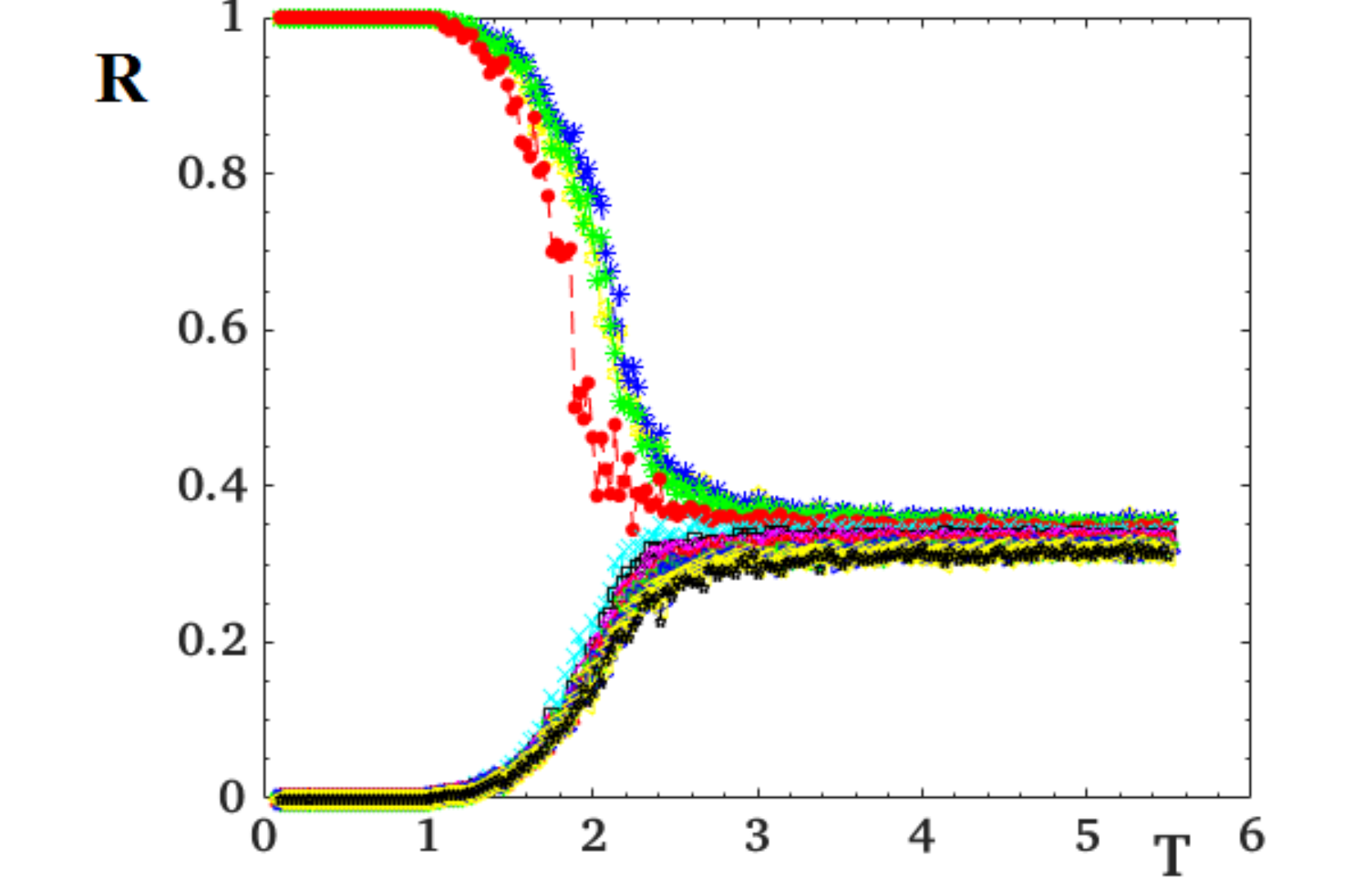}
\caption{Occupation rate $R$ per layer vs $T$.The lattice size of the system is $15 \times 15 \times 12$ with $J_{\para}=4$, $J_\perp = 1$, and $D=0.7$. The concentration is fixed to $c=30\%$, \emph{i.e.} 4 filled layers in the solid phase.  Each color represents a layer.}\label{rate}
\end{figure}

Finally, we show in Fig. \ref{rate} the occupation rate $R$ of each layer in the system in
the case $c=30$\%. In the GS, and at low $T (<1)$ only the first four layers are occupied ($R~1$), the other layers are empty ($R~0)$. However with increasing $T$, $R$ of the first four layers diminish and $R$ of the other layers increase. For very high $T$, the spins occupy all layers, and it is only  after the melting that all layers have the same occupation rate as expected in the liquid phase.

To close this section, let us discuss about the values of interaction used above. The value $J_{\perp}=1$ was used as the energy unit. The value $J_{\para}=4$ was used to favor layered ordering. As long as $J_{\para}>J_\perp$ our results shown above do not change qualitatively, only the range of values of $D$ giving rise to the vortex structure as well as the transition temperature change. Finally, let us note that the results have been shown for the same lateral lattice size but changing this size will not change qualitatively the results except the change of the range of values of $D$ giving rise to the vortex structure because of the long-range nature of the dipolar interaction.

\section{Conclusion}\label{Concl}
In this paper, we have studied a dot where the lattice sites are occupied by mobile Heisenberg spins. The dot is embedded in a close recipient which allows to conserve the number of spins. We have taken into account the  in-plane and perpendicular exchange interactions and the long-range dipolar interaction without cut-off. The confined geometry of the dot and the competition between exchange interactions and the dipolar interaction gives rise to a ground state which is a vortex around the dot center with the spins at the center pointing out of the $xy$ plane. Such a structure has a net perpendicular magnetization with a two-fold degeneracy along the $\pm z$ axis. Using a small magnetic field one can pin the magnetization in + or -$z$ direction. If the dot is sandwiched between two ferromagnetic films, then one can obtain a giant magneto-resistance \cite{Fert-Baibich,Grunberg} in a perpendicular spin transport.  Let us denote the up spin of the ferromagnetic film by $\uparrow_F$, and the up dot magnetization by $\uparrow_D$. According to the giant magneto-resistance geometry, the configuration  $\uparrow_F|\uparrow_D|\uparrow_F$   will let the electron up spins go across the system (high current) while $\uparrow_F|\downarrow_D|\uparrow_F$ will block the electron up spins (small current). The switch between the two states can be realized with an applied magnetic field. Thanks to the smallness of the dot magnetization, one just needs a small magnetic field which does not heat the system with the magnetization reversal. This is certainly an advantage over the use of a ferromagnetic layer instead of the dot.

Let us discuss another possible application of the present system in the domain of computer memory devices. The present system has a two-level structure which is stable at finite temperatures below the melting point (cf. Fig. \ref{EAD}): center spins can be up or down piloted by an extremely small magnetic field to reverse just a few spins as said earlier. This can serve as a two-bit unit. An application device one can imagine is a, array of dots with horizontal lines are the bit lines and vertical lines are the word lines, similarly to what proposed in Ref. \cite{Wang-etal-nature} using array of dots of skyrmions. The present system represents a considerable advantage: small dot size (just a few dozen of spins) and an extremely small magnetic field to operate. It can therefore increase the stocking capacity in memory devices for example.

We have studied the melting of such a dot with increasing temperature and found that, among other results, within the studied sizes the melting to the liquid phase takes place at the same temperature regardless of the system size. This is not the case of the order-disorder phase transition in a solid film \cite{DiepTM}.  Note that our technique using the mobile spin model can be used to study the behavior of liquid crystals. We have recently succeeded to obtain the nematic and smectic structures while cooling a liquid to low $T$ using an appropriate choice of Hamiltonian \cite{ABR2020,Ngo2020}.

Finally, note that in this work, we have simulated the system at a fixed concentration using the canonical description so that only one phase transition is observed. However, if we use the grand-canonical method, we believe that we will observe the coexisting phases at a given $T$ when varying the concentration, as what has been found in Ref. \cite{ABR-HTD-MK} using the mean-field approximation (cf. Fig. 6 of that reference). The implementation of the grand-canonical Monte Carlo method however is very complicated. This is left for a future study.


{}

\end{document}